# Weather impacts expressed sentiment

Patrick Baylis[1], Nick Obradovich[2,3] Yury Kryvasheyeu[4,5] Haohui Chen[6,7] Lorenzo Coviello[8], Esteban Moro[9], Manuel Cebrian[10], James H. Fowler[11]

We conduct the largest ever investigation into the relationship between meteorological conditions and the sentiment of human expressions. To do this, we employ over three and a half billion social media posts from tens of millions of individuals from both Facebook and Twitter between 2009 and 2016. We find that cold temperatures, hot temperatures, precipitation, narrower daily temperature ranges, humidity, and cloud cover are all associated with worsened expressions of sentiment, even when excluding weather-related posts. We compare the magnitude of our estimates with the effect sizes associated with notable historical events occurring within our data.

## Introduction

Mood and emotional state support human physical, psychological, and economic well-being. Positive emotions are associated with improved physiological factors such as cortisol levels and cardiovascular functioning[1] and amplify cognitive performance and mental flexibility[2]. They can also increase social connectedness and perceived social support[3] and may augment income and economic success[4]. Emotional states can also be transmitted through social networks[5,6], amplifying the broad-scale effects of altered individual emotions.

1 Center on Food Security and the Environment, Stanford University

2 Corresponding author, nobradov@mit.edu. P.B. and N.O. share equal contribution.

3 Media Lab, Massachusetts Institute of Technology

4 Data61, Commonwealth Scientific and Industrial Research Organisation (CSIRO), Australia

5 Faculty of Information Technology, Monash University

6 Data61, Commonwealth Scientific and Industrial Research Organisation (CSIRO), Australia

7 Faculty of Information Technology, Monash University

8 Institute of Electrical and Electronics Engineers

9 Department of Mathematics and GISC, Universidad Carlos III de Madrid

10 Data61, Commonwealth Scientific and Industrial Research Organisation (CSIRO), Australia

11 Departments of Political Science and Medicine, UC San Diego

Prior work suggests that environmental factors -- and ambient meteorological conditions in particular -- may substantively impact emotional state. However, previous empirical investigations of this relationship have found conflicting results. Early studies found large associations between meteorological conditions and mood[7,8] but were limited by small sample sizes. A number of studies in the most recent decade have found small to negligible associations[9–11], while others document associations that vary across individuals[12,13], associations observed at high levels of aggregation[14], or associations that are contingent on other factors[15]. A still more recent large-scale longitudinal analysis reports robust linkages between daily weather variation and reported well-being[16]. Whether, and if so, how meteorological variables shape human emotions remains an open question.

This pattern of divergent results is due in part to a lack of large-scale data on emotional states. To rectify this problem, we employ a correlate of emotional states: the sentiment of human lexical expressions on social media[17]. We report on associations between meteorological conditions and the expressed sentiment of tens of millions of United States residents across 3.5 billion posts on both Facebook and Twitter between 2009 and 2016. This work expands on Baylis (2015), who uses a billion Twitter posts over a shorter timeframe to estimate preferences for and valuations of different realizations of temperature in order to project the amenity cost of climate change[18]. In particular, the significantly longer sample period and the inclusion of the more representative, longer-form Facebook data in our analysis allows us to defensibly generalize our findings more broadly than was possible in Baylis (2015).

In this manuscript, we examine three questions. First, do weather conditions associate with changes in the sentiment of human expressions? Second, are these associations robust to excluding discussion about the weather itself? Third, how do the magnitudes of these weather-sentiment associations compare to the effect sizes of other events that alter expressed sentiment?

## Methods

## Data collection procedure

### Social media data

Our social media data consist of 3.5 billion posts in total, with 2.4 billion from Facebook and with 1.1 billion from Twitter. By using data from both social media platforms, we take advantage of the relative strengths of each as a data source on sentiment: Facebook data is more likely to be representative and to consist of text expressions revealing the user's underlying emotional state, while Twitter data allows additional investigation into the mechanisms underlying these changes in expressed sentiment and to compare the effect sizes to other events.

### Facebook data

To measure expression of sentiment on Facebook, we use data from a previous work[5]. These data are based on "status updates" which are text-based messages that a user's contacts may view on their own Facebook News Feed. Our Facebook dataset starts on January 1st, 2009 and ends on March 31st, 2012, with 1,176 days in total.

The Facebook data contain all users that chose English as their language, selected United States as their country, and could be matched to our selection of metropolitan areas by their IP-based geographic location. The Facebook data we use here are described in detail elsewhere[5]. A central benefit of using these data is that the Facebook population is more likely to reflect the population at large. By the end of our sampling period, nearly 70% of online adults used Facebook, as compared to between 15 and 30% for Twitter, LinkedIn, Pinterest, and Instagram[19]. Surveyed adults also indicated that they use Facebook more frequently than any other social media platform.

### Twitter Data

Our Twitter data consist of posts, or "tweets", that are short messages limited to 140 characters and are publicly viewable by default. Our Twitter data cover the period from November 30th, 2013 to June 30th, 2016, with 938 days in total.

We collected tweets using Twitter's public Streaming API, placing a bounding box filter over the United States to gather our sample of precisely geo-located tweets. We then assigned tweets falling within a metropolitan area's boundaries to that specific area. This procedure allows for a high level of certainty that the tweet originated within a specific metropolitan area.

## Meteorological data

We employ gridded meteorological data from the PRISM Climate Group for our daily maximum temperature, temperature range, and precipitation measures[20]. We also employ cloud cover and relative humidity data from the National Centers for Environmental Prediction (NCEP) Reanalysis II project[21] as these measures have been central to previous studies of the relationship between weather and emotional states[7,8].[12]

## Measures of expressed sentiment

To determine whether a social media post uses words that express positive or negative sentiment, we rely on the Linguistic Inquiry Word Count (LIWC) sentiment analysis tool[22]. LIWC is a highly validated, dictionary-based, sentiment classification tool that is commonly used to assess sentiment in social media posts[5,6,23,24] (of note, our results are similar under the use of alternative sentiment classifiers, see *SI:*

---

12 As daily NCEP data run only to June 30th 2016 at the time of writing, we use that date as the end date of our analysis.

*Alternative measures of expressed sentiment*). In our analysis, we treat positive and negative sentiment as separate constructs[25].

## Dependent variable aggregation

We aggregate both our Facebook and Twitter data to the 75 most populated metropolitan areas in the US. For each post, we calculate whether the post contains either a positive or a negative LIWC term. We then average each user's posts on each day to produce a positive rate and a negative rate for each user. Then, for each city on each day, we average the values of users in that city to calculate our city-level dependent variables. In order to ensure our results aren't driven by ecological inference, we present user-level analyses in *SI: User-level analysis*.

## Analyses

### City-level

To investigate if weather alters expressed sentiment at the city-level, we combine our aggregated city-level positive and negative sentiment scores with our daily meteorological data. We empirically model this relationship as:

$$Y_{jmt} = f(tmax_{jmt}) + g(precip_{jmt}) + h(\mu) + \alpha_j + \gamma_t + v_{jm} + \epsilon_{jmt} \quad (1)$$

In this time-series cross-sectional model, $j$ indexes cities, $m$ indexes unique year-months, and $t$ indexes calendar days. Our dependent variable $Y_{jmt}$ represents our city-level measure of positive or negative expressed sentiment, respectively.

Our independent variables of interest are maximum temperatures ( $tmax_{jmt}$ ) and total precipitation ( $precip_{jmt}$ ). We also examine the marginal effects of temperature range, percentage cloud cover, and relative humidity, represented via $h(\mu)$. We empirically estimate our relationships of interest using indicator variables for each 5℃ maximum temperature and temperature range bin, for each 1cm precipitation bin, and for each 20 percentage point bin of cloud cover and relative humidity (represented here by $f()$ , $g()$ , and $h()$ respectively). This procedure allows for flexible estimation of the association between our meteorological variables and expressed sentiment[26–28].

Unobserved geographic or temporal factors may influence sentiment in a way that correlates with weather. For example, people may be happier on average in cities that have better infrastructure or on days when they are likely to have more leisure time. To ensure that these factors do not bias our estimates of the association between weather and expressed sentiment, we include in Equation 1 $\alpha_i$ and $\gamma_t$ to represent city and calendar date indicator variables, respectively. These variables control for all constant unobserved characteristics for each city and for each day[29]. Further, there may be unobserved, city-specific trends, such as changes in amount of daylight throughout the year or evolution in city-level economic conditions over time, that influence the expressed sentiment of a city. In order to control for these potential confounds we include $v_{jm}$ in Equation 1, representing city-specific year-month indicator variables[30].

We adjust for within-city and within-day correlation in $\epsilon_{jmt}$ by employing heteroskedasticity-robust standard errors clustered on both city-year-month and day[31]. We exclude non-climatic control variables from Equation 1 because of their potential to generate bias in our parameters of interest[30,32]. Finally, we weight the city-level regression by the number of underlying social media posts for each city-day.

We omit the 20°C-25°C maximum temperature, the 0°C-5°C temperature range, 0cm precipitation, 0-20% cloud cover, and 40-60% humidity indicator variables when estimating Equation 1. We interpret our estimates as the percentage point change in positive or negative expressed sentiment associated with a particular meteorological observation range relative to these baseline categories.

## Exclusion of weather terms

Our first analyses examine the sentiment of all expressions contained within our data, inclusive of terms that may refer directly to the weather. As weather discussion may not necessarily reflect changes in individuals' underlying emotional states, however, we use a large dictionary of weather terms to filter out posts in our Twitter data that contain a plausible reference to the weather, and again run the models in Equation 1 and Equation 2 on the messages that do not contain these weather related terms.[13] Approximately 4% of tweets contained one or more of our weather terms. To investigate the effectiveness of this filter, we manually classified a random sample of 1,000 posts that included a weather term and determined that approximately 28% of that sample were about the weather. We also manually classified a second random sample of 1,000 non-weather term posts and determined that only 0.2% were about weather-related constructs. We present our weather-term dictionary in *SI: Weather terms*.

## Effect sizes in context

To contextualize our estimates, we compare the effect size of the association between below freezing temperatures and non-weather related positive expressed sentiment to the effect size of a number of plausibly negative events that occurred over the time span of our Twitter data. To do so, we again estimate both Equations 1 and 2. But, in addition to the terms in those models, we include indicator terms for each of the comparison events to estimate these parameters simultaneously alongside our meteorological variables. These indicator terms isolate the specific location and the specific dates of the event so that they are not collinear with the fixed effects in our models. For example, for the effect size of the San Francisco Bay

---

13 Because we no longer have access to the raw Facebook posts, we were unable to re-calculate our non-weather expressed sentiment metrics for our Facebook corpus.

Area earthquake, we create an indicator variable for the San Francisco/Oakland metropolitan area on August 24th 2014, the date of the earthquake.

# Results

## Descriptive statistics

***Table 1.*** *Summary statistics of main dependent and independent variables.*

|  | Facebook City Mean | Facebook City Std. Dev. | Twitter City Mean | Twitter City Std. Dev. |
|---|---|---|---|---|
| Pos. Rate | 41 | 4.39 | 34.93 | 2.72 |
| Neg. Rate | 21.4 | 3.25 | 18.54 | 2.43 |
| Pos. No Wth. | NA | NA | 34.55 | 2.71 |
| Neg. No Wth. | NA | NA | 18.58 | 2.46 |
| Max. Temperature | 20.38 | 10.79 | 21.24 | 10.83 |
| Precipitation | 0.25 | 0.78 | 0.26 | 0.82 |
| Cloud Cover | 36.5 | 27.41 | 40.09 | 27.09 |
| Humidity | 68.23 | 18.52 | 67.64 | 18 |

We present the descriptive statistics associated with our main variables in Table 1.

## All expressed sentiment

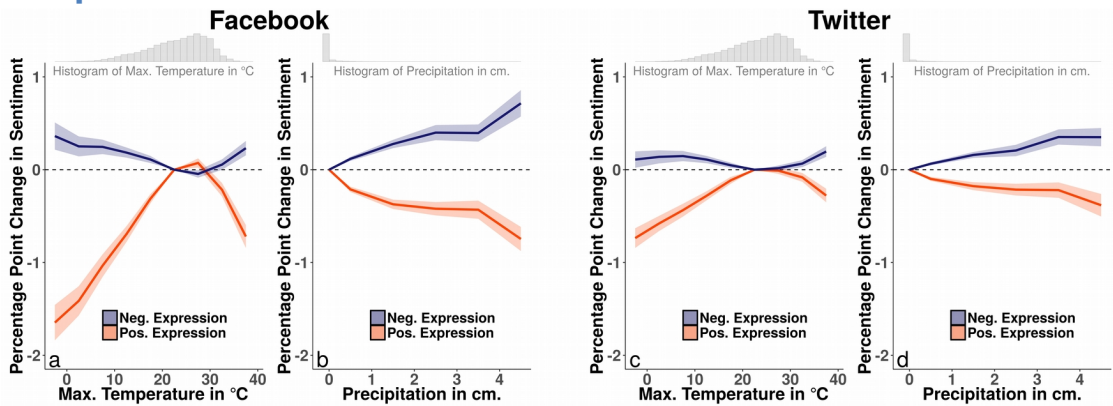

***Fig. 1.*** *Facebook and Twitter analyses for all message types. Panel (a) depicts the relationship between daily maximum temperatures and the rates of expressed sentiment of approximately 2.4 billion Facebook status updates from 2009-2012, aggregated to the city-level. It draws from the estimation of Equation 1 and plots the predicted change in expressed sentiment associated with each maximum temperature*

*bin. Panel (b) depicts the relationship between daily precipitation and the rates of sentiment expression of Facebook status updates, also drawing on estimation of Equation 1. Panels (c) and (d) replicate these analyses for nearly 1.1 billion Twitter posts between 2013 and 2016 aggregated to the same cities. Shaded error bounds represent 95% confidence intervals.*

The results of estimating Equation 1 for the association between meteorological conditions and positive and negative sentiment on Facebook indicate that temperature, precipitation, humidity, and cloud cover each significantly relate to expressions of sentiment (see Figure 1 panels (a) and (b) for the maximum temperature and precipitation results). Positive expressions increase up to maximum temperatures of 20°C and decline past 30°C. The impact of temperature on negative expressions is qualitatively the opposite of its impact on positive expressions, though smaller in magnitude. Precipitation worsens expressed sentiment. Daily temperature ranges exceeding 15°C are associated with significant increases in positive expressions (coefficient: 0.104, p: 0.002, n: 85,801) and reductions in negative expressions (coefficient: -0.177, p: < 0.001, n: 85,801). Levels of relative humidity exceeding 80% decrease positive expressions (coefficient: -0.084, p: 0.001, n: 85,801) and increase negative expressions (coefficient: 0.039, p: 0.038, n: 85,801), as do days with high cloud cover, reducing positive expressions (coefficient: -0.2, p: < 0.001, n: 85,801) and increasing negative ones (coefficient: 0.168, p: < 0.001, n: 85,801) (see *SI: Regression tables* for further details on these results).

Panels (c) and (d) of Figure 1 display the results of estimating Equation 1 on the Twitter city-level data. The nature of the impact of temperature and precipitation on sentiment expression is quite similar to the associations in the Facebook data, though attenuated in magnitude. For example, the association between below freezing temperatures and city-level positive sentiment expressions on Twitter is approximately 45% the size of this parameter in the Facebook data. The effect sizes of precipitation, temperature range, cloud cover, and humidity on expressed sentiment on city-level Twitter data retain statistical significance but are similarly attenuated in magnitude as compared to Facebook (see *SI: Regression tables* for more details).

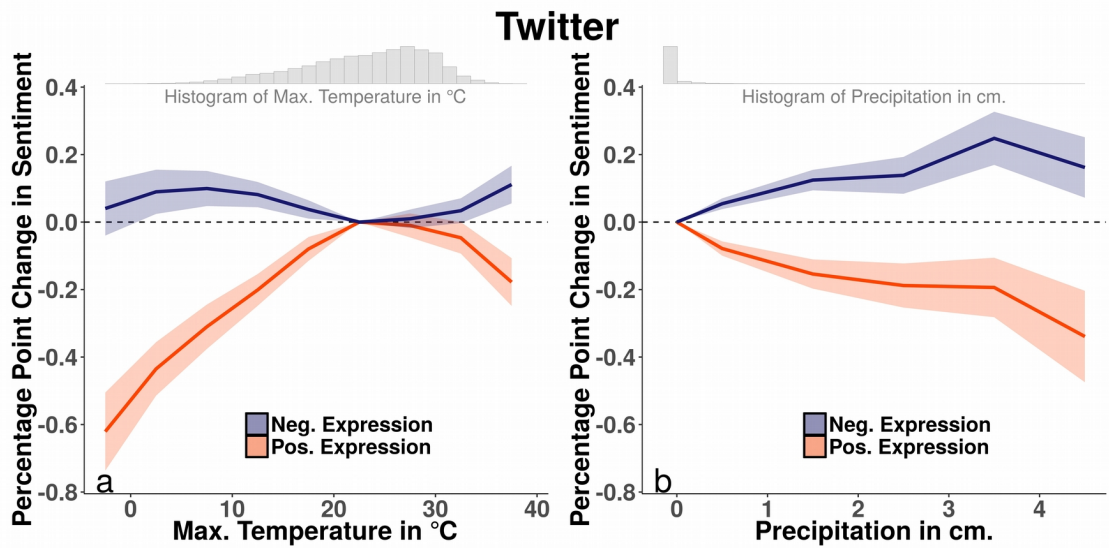

**Fig. 2.** *Twitter analyses for posts without weather terms. Panel (a) depicts the relationship between daily maximum temperatures and the rates of expressed sentiment for non-weather posts, aggregated to the city-level. It draws from the estimation of Equation 1 and plots the predicted change in expressed sentiment associated with each maximum temperature bin. Panel (b) depicts the relationship between daily precipitation and the rates of sentiment expression of non-weather posts, also drawing on estimation of Equation 1. Shaded error bounds represent 95% confidence intervals.*

Analyzing all posts provides useful descriptive characteristics of the ways in which aggregate sentiment associates with the weather. However, in order to employ expressed sentiment as a better proxy for underlying emotional states, it is useful to exclude direct references to the weather itself, as more extreme weather conditions significantly alter both the rate of weather discussion and the sentiment of this discussion (see *SI: Rate of messages containing weather terms* and *SI: Sentiment of messages containing weather terms*). We present the results of estimating Equation 1 on our corpus of Twitter posts excluding weather terms in Figure 2.

Panels (a) and (b) of Figure 2, illustrate that the effect sizes associated with maximum temperature and precipitation on non-weather sentiment at the city-level are slightly smaller than the effect sizes seen in the all-posts sample (Figure 1 panels (c) and (d)). As an example, the association between below freezing temperatures and positive sentiment, non-weather-term expressions is approximately 84% the size of this association in the all-posts model. The effect sizes of temperature range, high humidity, and cloud cover retain significance in this model, though are also attenuated in size compared to the all-posts model (see *SI: Regression tables* for details).

## Effect sizes in context

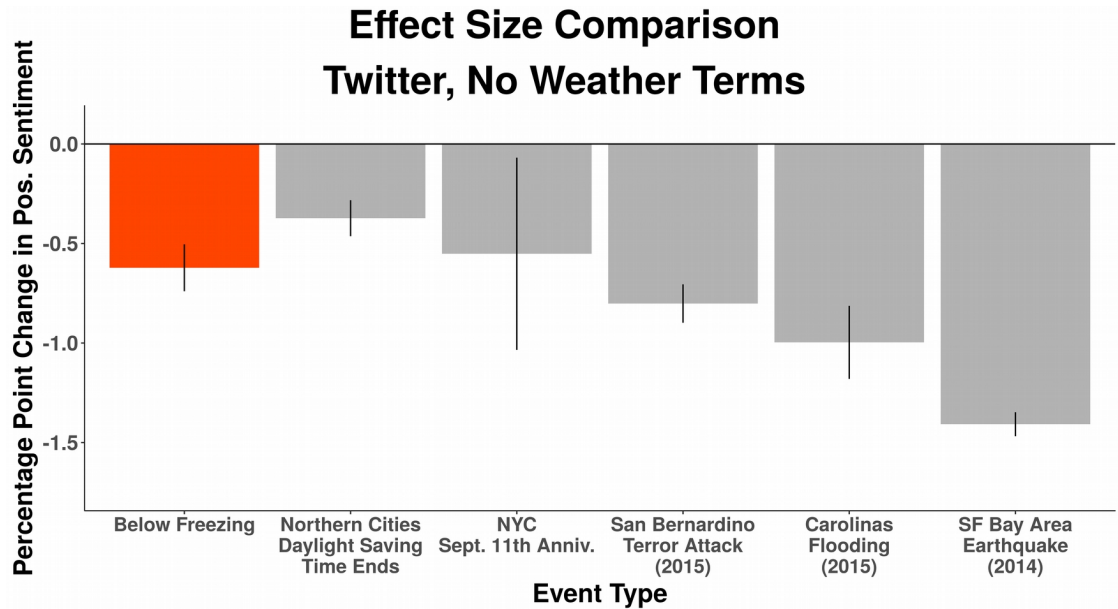

**Fig. 3.** *Comparisons between the effect size of below freezing temperatures on positive, non-weather, expressed sentiment with the effect sizes of other locale-specific events over the course of our data on the same sentiment metric at the Twitter city-level. The effect size of freezing temperatures compares in magnitude to other significant events.*

To understand the relative magnitude of these predicted changes, it can be helpful to look at the effect sizes associated with other types of events on expressed sentiment. We chose five salient events for their theoretically negative association with expressed positive, non-weather sentiment coupled with their diversity of type: 1) the end of daylight saving time in the cities above the median latitude of our sample of cities 2) the annual anniversary of the September 11th terrorist attacks in the New York City metropolitan area 3) the December 2015 terrorist attacks in the San Bernardino metropolitan area 4) the October 2015 flooding in the Carolinas in the Charlotte metropolitan area[14] and 5) the San Francisco Bay Area earthquake in August 2014 in the San Francisco/Oakland metropolitan area.

Figure 3 indicates that each of these events is significantly associated with reductions in non-weather, expressed positive sentiments in the areas local to the events. Further, a day of below freezing temperature in our sample is substantively meaningful. For example, the effect size of a day of below freezing temperature on positive expressed sentiment is 62% the effect size of the 2015 Carolina floods on aggregate expressed sentiment in Charlotte.[15]

---

14 The term 'flood' and its derivatives are part of our excluded weather terms.

## Discussion

There are several considerations important to the interpretation of our results. First, while we have data on millions of individuals' expressed sentiment as reflected by their social media posts, optimal data would also include these individuals' daily self-reported emotional states. As mentioned above, while sentiment expressions on social media can be reflective of underlying emotions[17], the linguistic measures we employ here represent an imperfect and noisy proxy of emotional states. Future studies are needed to improve the psychometric validity of sentiment metrics.

Second, and relatedly, our chosen LIWC sentiment metrics may imperfectly measure the sentiment of expressions on social media. We examine the robustness of our findings to the use of other sentiment classification tools with our Twitter data in *SI: Alternative measures of expressed sentiment*. In those analyses we employ both the SentiStrength and Hedonometer algorithms and find that our results are quite robust across all three of our employed sentiment metrics[33,34]. However, because all three of our metrics likely have idiosyncratic errors associated with them, our measurement of the sentiment of expressions remains imperfect.

Third, measurement error may exist between observed weather and the weather users actually experience, possibly attenuating the magnitude of our estimates[35]. Issues of right-hand-side measurement error may be particularly salient with respect to our measures of cloud cover and humidity, as they are derived from reanalysis data rather than directly from station observations[30].

Fourth, our analysis is conducted on individuals who self-select into participation in social media. Our results may not apply to demographics that do not use either Facebook or Twitter, such as older generations. Because the elderly are less common users of social media and because they may be particularly vulnerable to adverse weather conditions[32], the results we present here may underestimate the true population associations between weather and expressed sentiment.

Fifth, our data are restricted to observations from one country with a predominately temperate climate and with one of the highest rates of air conditioning prevalence in the world[36]. It is critical to repeat this analysis where possible in poorer countries with more extreme climates as they may see even greater alterations in overall expressed sentiment due to meteorological conditions.

---

15 Of important note, by conducting this comparison, we do not intend to comment on the overall impact of these events on society or on individual well-being. There may be many reasons why, for example, the San Bernardino shootings produced a less substantial reduction in positive expressed sentiment on Twitter than did the SF Bay Area earthquake. Speculation on the underlying factors that drive the association between these specific events and expressed sentiment is beyond the scope of this study.

Ultimately, given the ubiquity of our exposure to varying weather conditions, understanding the influence they may have on our emotional states is of high importance. Here we provide a window into this relationship via the measurement of expressed sentiment on social media. We observe that the weather is associated with statistically significant and substantively meaningful changes in expressed sentiment for posts both inclusive of and exclusive of weather-related terms. We find substantial evidence that less ideal weather conditions relate to worsened sentiment. To the extent that the sentiment of expressions serves as a valid proxy for underlying emotions, we find some observational evidence that the weather may functionally alter human emotional states.

## Author Contributions

P.B. and N.O. conceived of the research question and developed figures. N.O. and P.B. analyzed the data. N.O., H.C., Y.K., and L.C. constructed the historical data. N.O. drafted the manuscript and compiled supplementary information. All authors edited the manuscript and supplementary information and aided in concept development.

## Acknowledgements

We thank the San Diego Supercomputer Center, Zachary Steinert-Threlkeld, Jason J. Jones for their assistance and thank Uri Simonsohn, Tal Yarkoni, and our anonymous reviewers for their thorough and helpful feedback.

## Competing Financial Interests Statement

The authors declare that they have no conflicts of interest with respect to their authorship or the publication of this article.

## Data Availability

We collected our Twitter data from the public domain in adherence with Twitter's Developer Agreement and we used aggregated Facebook data published previously[5]. The social media data used in this study are restricted from public redistribution.

## Funding

This work was supported by the National Science Foundation (Grant Nos. DGE0707423, TG-SES130013, and 0903551 to N.O.) and by the Ministerio de Economia y Competitividad (Spain) (Grants FIS2013-47532-C3-3-P and FIS2016-78904-C3-3-P to E.M.).

# Weather impacts expressed sentiment

## Supplementary Information

*Patrick Baylis, Nick Obradovich, Yury Kryvasheyeu, Haohui Chen, Lorenzo Coviello,
Esteban Moro, Manuel Cebrian, James H. Fowler*

# Contents

# User-level analysis

## Methods

For our user-level analysis, we employ the posts of users who authored Twitter messages on greater than 25% of days in our sample, a subsample containing 511 million tweets across 365,476 users. We calculate our expressed sentiment dependent variables identically to the city-level analysis, stopping at the user-level of aggregation. For each user-day we have the percentage of that user's tweets that contain positive sentiment as well as the percentage of the user's tweets that contain negative sentiment. We restrict our user-level analysis to our Twitter data as we no longer retain access to the user-level Facebook data.

To investigate if the weather is associated with alterations of expressed sentiments within individuals over time, we employ our user-level data, along with a slightly modified equation from the main text. We model our user-level relationship as:

$$Y_{ijmt} = f(tmax_{ijmt}) + g(precip_{ijmt}) + h(\mu) + \eta_i + \gamma_t + \nu_{jm} + \epsilon_{ijmt} \qquad (S1)$$

In Equation S1 $i$ now indexes unique individuals and $\eta_i$ replaces $\alpha_j$ and represents user-level indicator terms that control for individual-specific, time-invariant factors such as average mood, constant demographic covariates, and fixed weather preferences[1]. The model again includes calendar date and city-level by year-month indicator terms.

## Descriptive statistics

Table 1: Summary statistics of main dependent and independent variables.

|                  | Twitter User Mean | Twitter User Std. Dev. |
|------------------|-------------------|------------------------|
| Pos. Rate        | 33.32             | 35.02                  |
| Neg. Rate        | 18.13             | 27.52                  |
| Pos. No Wth.     | 32.99             | 35.06                  |
| Neg. No Wth.     | 18.1              | 27.64                  |
| Max. Temperature | 21.69             | 10.32                  |
| Precipitation    | 0.27              | 0.83                   |
| Cloud Cover      | 39.8              | 26.88                  |
| Humidity         | 68.39             | 16.7                   |

We present the descriptive statistics associated with our main variables in Table S1.

## All expressed sentiment

Panels (a) and (b) of Figure S1 display the results of estimating Equation S1 on 81,388,085 user-days of Twitter user-level data. The nature of the impact of temperature and precipitation on sentiment expression is quite similar to the effect size in the city-level data, though these are again attenuated in magnitude. The effect sizes of precipitation, temperature range, and cloud cover on user-level expressed sentiment on Twitter retain statistical significance but are similarly attenuated in magnitude as compared to the city-level Twitter model (see *Tables for Figure 1* for details). The association between high levels of humidity and positive expressed sentiment fails to gain significance in this model.

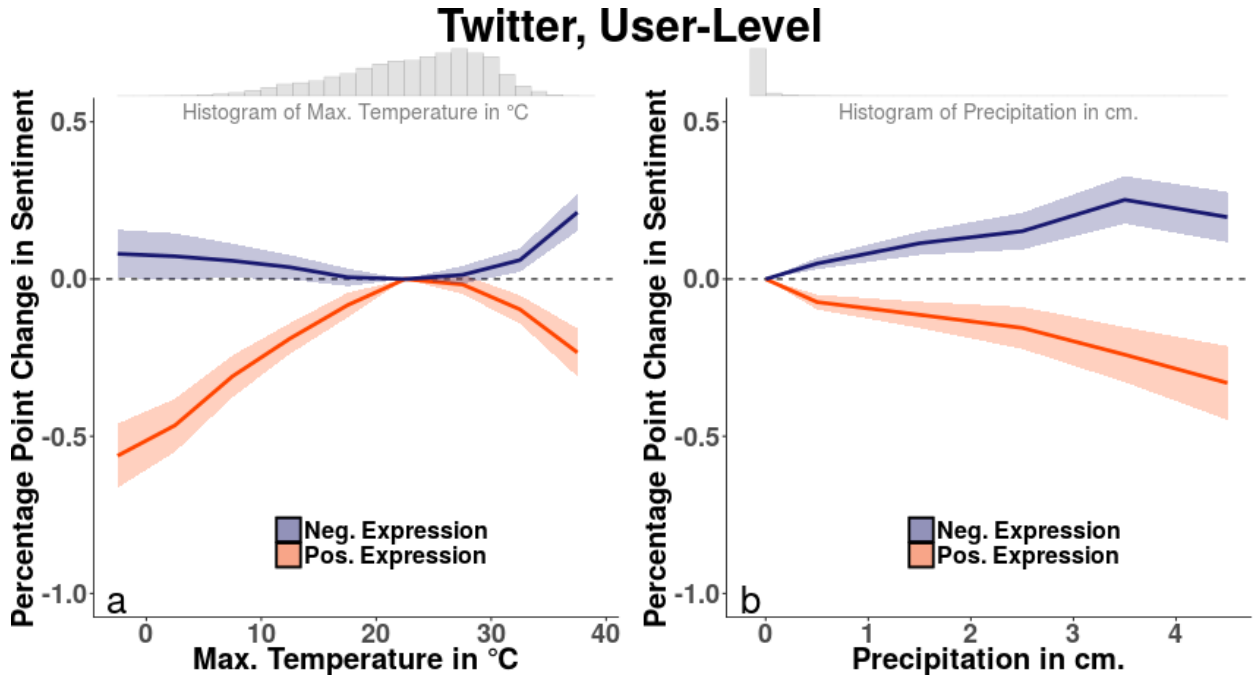

Figure 1: Twitter user-level analyses for all message types. Panels (a) and (b) replicate the main text analyses for Twitter user-level data. Shaded error bounds represent 95% confidence intervals.

## No weather terms

Excluding weather terms from the user-level data also reduces effect sizes somewhat, as can be seen for temperature and precipitation in panels (a) and (b) of Figure S2. In this model, the association between cold temperatures and large amounts of precipitation and negative expressed sentiment fail to gain statistical significance (though more moderate amounts of precipitation still significantly associate with increased rates of negative sentiment). High temperature ranges in this model still associate with improved expressed sentiment while cloud cover again associates with worsened sentiment. Humidity fails to gain significance, though the signs of the associations remain the same (see *Tables for Figure 2* for further details).

## Effect sizes in context

At the user level, the effect sizes associated with the weather are smaller than at the city-level but are still meaningful (see Figure S3). At this level, a day of below freezing temperature is associated with 37% the effect size of the Carolina flooding on user-level expressed sentiment in Charlotte.

## Weather terms

Below we list the 318 crowd-sourced weather terms that we exclude at the tweet level in order to calculate our non-weather related sentiment metrics.

aerovane air airstream altocumulus altostratus anemometer anemometers anticyclone anticyclones arctic arid aridity atmosphere atmospheric autumn autumnal balmy baroclinic barometer barometers barometric blizzard blizzards blustering blustery breeze breezes breezy brisk calm celsius chill chilled chillier chilliest chilly chinook cirrocumulus cirrostratus cirrus climate climates cloud cloudburst cloudbursts cloudier cloudiest clouds cloudy cold colder coldest condensation

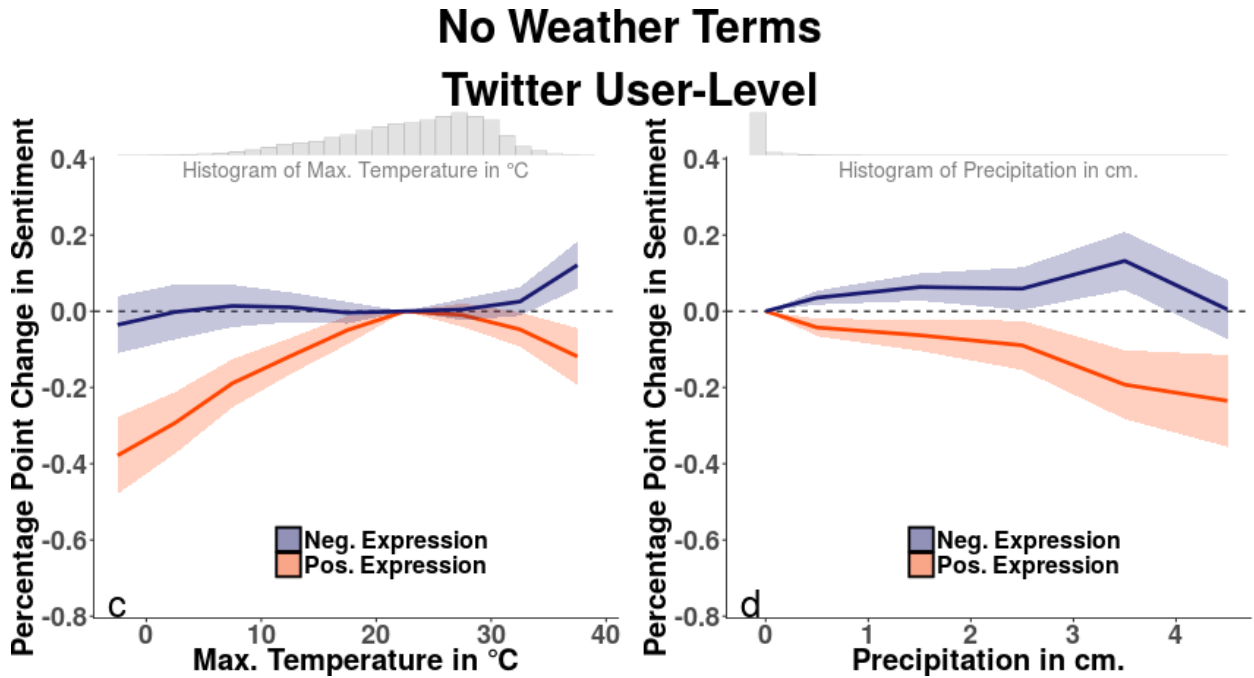

Figure 2: Twitter user-level analyses for posts without weather terms. Panels (a) and (b) depict the results of estimating Equation 2 on the sentiment of non-weather posts at the user level. Shaded error bounds represent 95% confidence intervals.

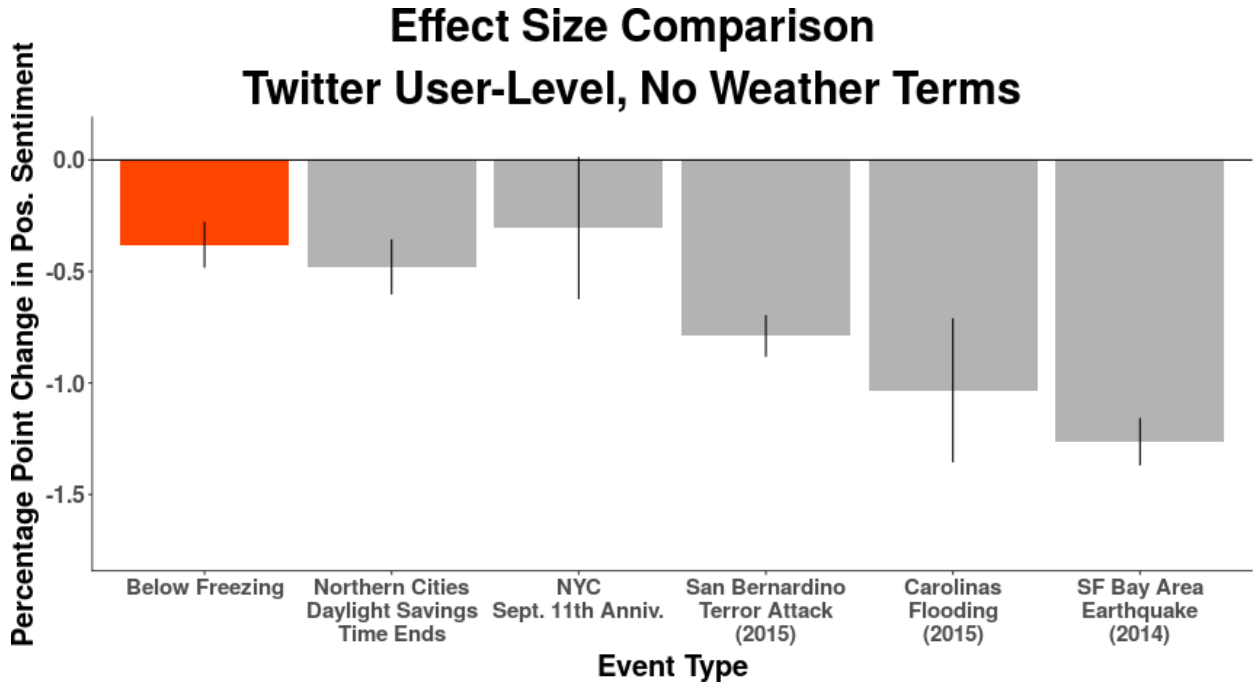

Figure 3: User-level comparisons between the effect size of below freezing temperatures on positive, non-weather, expressed sentiment with the effect sizes of other locale-specific events over the course of our data on the same sentiment metric. The effect size of freezing temperatures compares in magnitude to other significant events.

contrail contrails cool cooled cooling cools cumulonimbus cumulus cyclone cyclones damp damp
damper damper dampest dampest degree degrees deluge dew dews dewy doppler downburst
downbursts downdraft downdrafts downpour downpours dried drier dries driest drizzle drizzled
drizzles drizzly drought droughts dry dryline fall farenheit flood flooded flooding floods flurries
flurry fog fogbow fogbows fogged fogging foggy fogs forecast forecasted forecasting forecasts freeze
freezes freezing frigid frost frostier frostiest frosts frosty froze frozen gale gales galoshes gust
gusting gusts gusty haboob haboobs hail hailed hailing hails haze hazes hazy heat heated heating
heats hoarfrost hot hotter hottest humid humidity hurricane hurricanes ice iced ices icing icy
inclement landspout landspouts lightning lightnings macroburst macrobursts maelstrom mer-
cury meteorologic meteorologist meteorologists meteorology microburst microbursts microclimate
microclimates millibar millibars mist misted mists misty moist moisture monsoon monsoons
mugginess muggy nexrad nippy NOAA nor'easter nor'easters noreaster noreasters overcast ozone
parched parching pollen precipitate precipitated precipitates precipitating precipitation psychrom-
eter radar rain rainboots rainbow rainbows raincoat raincoats rained rainfall rainier rainiest
raining rains rainy sandstorm sandstorms scorcher scorching searing shower showering showers
skiff sleet slicker slickers slush slushy smog smoggier smoggiest smoggy snow snowed snowier
snowiest snowing snowmageddon snowpocalypse snows snowy spring sprinkle sprinkles sprinkling
squall squalls squally storm stormed stormier stormiest storming storms stormy stratocumulus
stratus subtropical summer summery sun sunnier sunniest sunny temperate temperature tempest
thaw thawed thawing thaws thermometer thunder thundered thundering thunders thunderstorm
thunderstorms tornadic tornado tornadoes tropical troposphere tsunami turbulent twister twisters
typhoon typhoons umbrella umbrellas vane warm warmed warming warms warmth waterspout
waterspouts weather wet wetter wettest wind windchill windchills windier windiest windspeed
windy winter wintery wintry

## Rate of messages containing weather terms

Below we analyze the effect of meteorological conditions on the rate of expressions that contain weather terms.
As can be seen in Figure S4, city-level rates and individual-level probabilities of weather speech, unsurprisingly,
increase and comprise a larger percentage of overall expressions under less pleasant meteorological conditions.

## Sentiment of messages containing weather terms

We also analyze the effect of meteorological conditions on the sentiment expressions that contain weather
terms. As can be seen in Figure S5, again unsurprisingly, both city-level and individual-level expressed
sentiment of weather messages markedly worsens under less pleasant meteorological conditions.

## Regression tables

### All posts

Table S2 corresponds to the city-level results associated with Equation 1 and Figure 1 from the main text.
Table S3 corresponds to the user-level results presented above.

### Posts without weather terms

Table S4 corresponds to the city-level results associated with Equation 1 and Figure 2 from the main text.
Table S5 corresponds to the user-level results presented above.

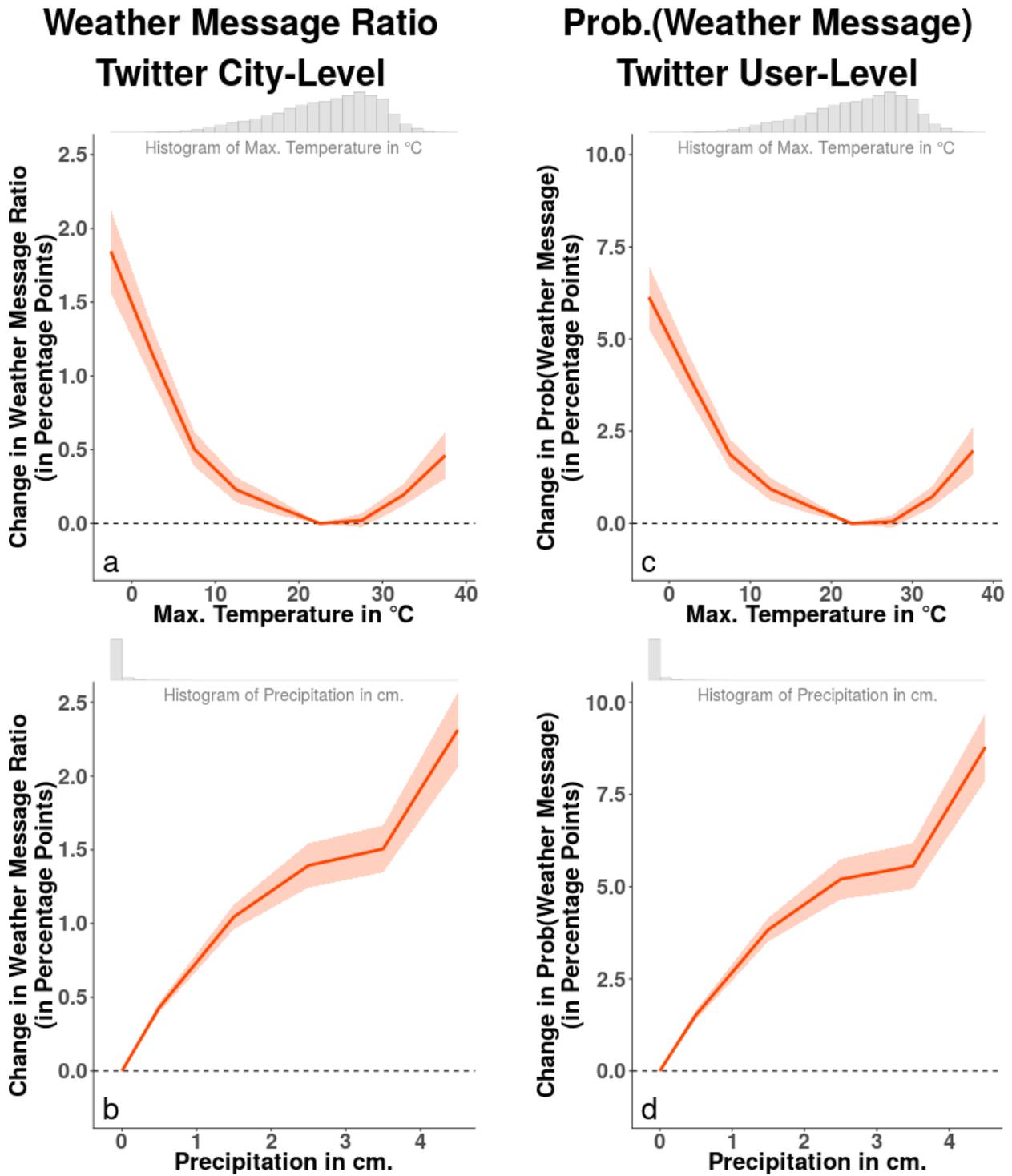

Figure 4: **Effects of the weather on city-level frequency and user-level probability of weather speech.**

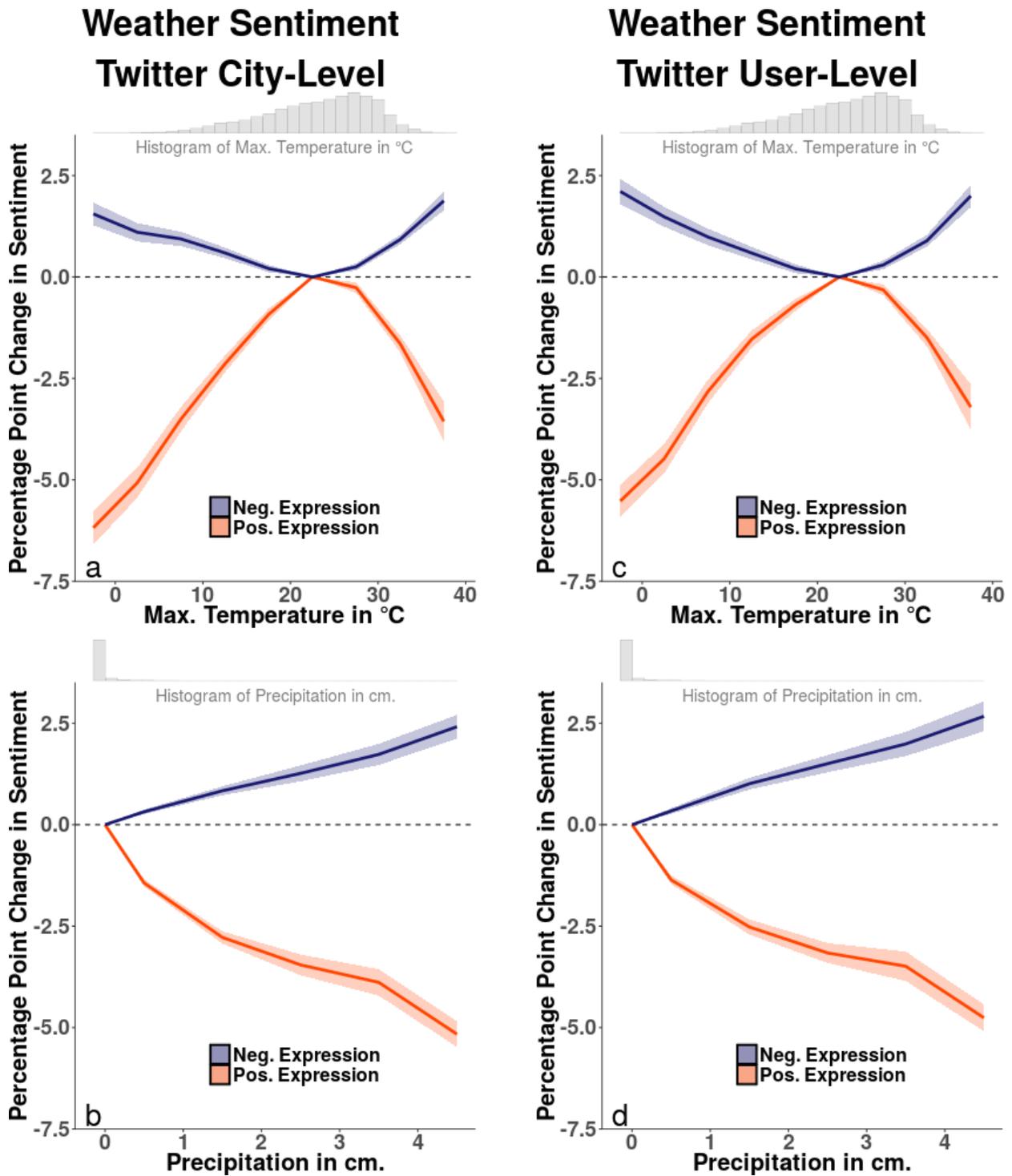

Figure 5: **Effects of the weather on expressed sentiment of posts that contain weather terms.**

Table 2: City-Level Weather and Expressed Sentiment, All Posts

| | *Dependent variable:* | | | |
|---|---|---|---|---|
| | Positive | Negative | Positive | Negative |
| | | Facebook | | Twitter |
| | (1) | (2) | (3) | (4) |
| cuttmax(-Inf,0] | −1.647*** | 0.362*** | −0.737*** | 0.109** |
| | (0.097) | (0.075) | (0.054) | (0.045) |
| cuttmax(0,5] | −1.413*** | 0.251*** | −0.580*** | 0.138*** |
| | (0.081) | (0.055) | (0.042) | (0.036) |
| cuttmax(5,10] | −1.034*** | 0.246*** | −0.436*** | 0.147*** |
| | (0.062) | (0.040) | (0.035) | (0.028) |
| cuttmax(10,15] | −0.693*** | 0.184*** | −0.280*** | 0.109*** |
| | (0.043) | (0.028) | (0.025) | (0.020) |
| cuttmax(15,20] | −0.313*** | 0.110*** | −0.115*** | 0.048*** |
| | (0.025) | (0.020) | (0.019) | (0.014) |
| cuttmax(25,30] | 0.073*** | −0.046** | −0.010 | 0.016 |
| | (0.025) | (0.020) | (0.018) | (0.015) |
| cuttmax(30,35] | −0.216*** | 0.055* | −0.083*** | 0.066*** |
| | (0.037) | (0.029) | (0.024) | (0.019) |
| cuttmax(35, Inf] | −0.720*** | 0.232*** | −0.278*** | 0.195*** |
| | (0.063) | (0.039) | (0.037) | (0.029) |
| cutprcp(0,1] | −0.212*** | 0.117*** | −0.099*** | 0.064*** |
| | (0.015) | (0.011) | (0.011) | (0.008) |
| cutprcp(1,2] | −0.373*** | 0.277*** | −0.177*** | 0.159*** |
| | (0.026) | (0.023) | (0.022) | (0.016) |
| cutprcp(2,3] | −0.421*** | 0.399*** | −0.215*** | 0.207*** |
| | (0.037) | (0.042) | (0.032) | (0.031) |
| cutprcp(3,4] | −0.431*** | 0.395*** | −0.220*** | 0.352*** |
| | (0.050) | (0.047) | (0.042) | (0.042) |
| cutprcp(4, Inf] | −0.747*** | 0.715*** | −0.383*** | 0.350*** |
| | (0.066) | (0.073) | (0.062) | (0.050) |
| cuttrange(5,10] | −0.034 | −0.076*** | 0.009 | −0.026* |
| | (0.023) | (0.020) | (0.020) | (0.015) |
| cuttrange(10,15] | 0.043 | −0.165*** | 0.034 | −0.072*** |
| | (0.027) | (0.022) | (0.023) | (0.018) |
| cuttrange(15, Inf] | 0.104*** | −0.177*** | 0.063** | −0.072*** |
| | (0.034) | (0.029) | (0.029) | (0.023) |
| cuthumid(-Inf,40] | −0.058* | 0.038 | −0.043* | 0.031 |
| | (0.031) | (0.026) | (0.026) | (0.020) |
| cuthumid(60,80] | −0.021 | −0.005 | −0.020 | −0.015 |
| | (0.019) | (0.015) | (0.014) | (0.010) |
| cuthumid(80, Inf] | −0.084*** | 0.039** | −0.050*** | 0.031** |
| | (0.024) | (0.019) | (0.018) | (0.013) |
| cutcloud(20,40] | −0.052*** | 0.021* | −0.034*** | 0.007 |
| | (0.014) | (0.012) | (0.012) | (0.010) |
| cutcloud(40,60] | −0.086*** | 0.049*** | −0.067*** | 0.036*** |
| | (0.018) | (0.014) | (0.014) | (0.011) |
| cutcloud(60,80] | −0.142*** | 0.105*** | −0.077*** | 0.057*** |
| | (0.021) | (0.018) | (0.016) | (0.013) |
| cutcloud(80, Inf] | −0.200*** | 0.168*** | −0.107*** | 0.077*** |
| | (0.028) | (0.024) | (0.020) | (0.017) |
| City FE | Yes | Yes | Yes | Yes |
| Date FE | Yes | Yes | Yes | Yes |
| City:Year-Month FE | Yes | Yes | Yes | Yes |
| Observations | 85,801 | 85,801 | 67,972 | 67,972 |
| $R^2$ | 0.954 | 0.942 | 0.883 | 0.902 |
| Adjusted $R^2$ | 0.951 | 0.939 | 0.877 | 0.897 |
| Residual Std. Error | 135.818 | 110.274 | 100.463 | 78.829 |

*Note:* 

*p<0.1; **p<0.05; ***p<0.01

Standard errors are in parentheses and are clustered on city-yearmonth and date.

Table 3: User-Level Weather and Expressed Sentiment, Twitter All Posts

| | *Dependent variable:* | |
| --- | --- | --- |
| | Positive | Negative |
| | (1) | (2) |
| cuttmax(-Inf,0] | −0.562*** | 0.080** |
| | (0.052) | (0.039) |
| cuttmax(0,5] | −0.465*** | 0.072* |
| | (0.043) | (0.037) |
| cuttmax(5,10] | −0.309*** | 0.058** |
| | (0.034) | (0.028) |
| cuttmax(10,15] | −0.189*** | 0.038* |
| | (0.025) | (0.020) |
| cuttmax(15,20] | −0.083*** | 0.006 |
| | (0.020) | (0.014) |
| cuttmax(25,30] | −0.017 | 0.013 |
| | (0.016) | (0.015) |
| cuttmax(30,35] | −0.096*** | 0.060*** |
| | (0.023) | (0.019) |
| cuttmax(35, Inf] | −0.234*** | 0.212*** |
| | (0.039) | (0.030) |
| cutprcp(0,1] | −0.073*** | 0.049*** |
| | (0.012) | (0.009) |
| cutprcp(1,2] | −0.114*** | 0.114*** |
| | (0.021) | (0.019) |
| cutprcp(2,3] | −0.155*** | 0.152*** |
| | (0.034) | (0.030) |
| cutprcp(3,4] | −0.240*** | 0.252*** |
| | (0.045) | (0.039) |
| cutprcp(4, Inf] | −0.330*** | 0.197*** |
| | (0.060) | (0.041) |
| cuttrange(5,10] | 0.025 | −0.035** |
| | (0.019) | (0.016) |
| cuttrange(10,15] | 0.057*** | −0.066*** |
| | (0.021) | (0.019) |
| cuttrange(15, Inf] | 0.094*** | −0.090*** |
| | (0.028) | (0.024) |
| cuthumid(-Inf,40] | −0.027 | 0.030 |
| | (0.027) | (0.024) |
| cuthumid(60,80] | 0.006 | −0.008 |
| | (0.013) | (0.011) |
| cuthumid(80, Inf] | −0.016 | 0.028** |
| | (0.017) | (0.014) |
| cutcloud(20,40] | −0.052*** | 0.010 |
| | (0.013) | (0.011) |
| cutcloud(40,60] | −0.072*** | 0.030** |
| | (0.014) | (0.012) |
| cutcloud(60,80] | −0.086*** | 0.035** |
| | (0.017) | (0.014) |
| cutcloud(80, Inf] | −0.116*** | 0.040** |
| | (0.021) | (0.017) |
| User FE | Yes | Yes |
| Date FE | Yes | Yes |
| City:Year-Month FE | Yes | Yes |
| Observations | 81,388,085 | 81,388,085 |
| $R^2$ | 0.129 | 0.110 |
| Adjusted $R^2$ | 0.125 | 0.105 |
| Residual Std. Error | 32.702 | 25.976 |

*Note:* *p<0.1; **p<0.05; ***p<0.01
Standard errors are in parentheses and are clustered on city-yearmonth and date.

Table 4: City-Level Weather and Expressed Sentiment, Twitter No Weather Terms

| | *Dependent variable:* | |
| --- | --- | --- |
| | Positive | Negative |
| | (1) | (2) |
| cuttmax(-Inf,0] | −0.621*** | 0.040 |
| | (0.059) | (0.041) |
| cuttmax(0,5] | −0.435*** | 0.090*** |
| | (0.040) | (0.033) |
| cuttmax(5,10] | −0.310*** | 0.099*** |
| | (0.034) | (0.027) |
| cuttmax(10,15] | −0.201*** | 0.081*** |
| | (0.024) | (0.019) |
| cuttmax(15,20] | −0.080*** | 0.037*** |
| | (0.018) | (0.014) |
| cuttmax(25,30] | −0.010 | 0.010 |
| | (0.018) | (0.014) |
| cuttmax(30,35] | −0.047* | 0.034* |
| | (0.024) | (0.019) |
| cuttmax(35, Inf] | −0.178*** | 0.111*** |
| | (0.036) | (0.028) |
| cutprcp(0,1] | −0.078*** | 0.054*** |
| | (0.011) | (0.008) |
| cutprcp(1,2] | −0.154*** | 0.124*** |
| | (0.022) | (0.015) |
| cutprcp(2,3] | −0.188*** | 0.139*** |
| | (0.033) | (0.028) |
| cutprcp(3,4] | −0.194*** | 0.248*** |
| | (0.045) | (0.040) |
| cutprcp(4, Inf] | −0.339*** | 0.162*** |
| | (0.069) | (0.046) |
| cuttrange(5,10] | 0.020 | −0.028* |
| | (0.020) | (0.015) |
| cuttrange(10,15] | 0.033 | −0.076*** |
| | (0.023) | (0.017) |
| cuttrange(15, Inf] | 0.052* | −0.082*** |
| | (0.029) | (0.022) |
| cuthumid(-Inf,40] | −0.031 | 0.024 |
| | (0.025) | (0.019) |
| cuthumid(60,80] | −0.028** | −0.012 |
| | (0.014) | (0.010) |
| cuthumid(80, Inf] | −0.055*** | 0.025* |
| | (0.018) | (0.013) |
| cutcloud(20,40] | −0.025** | 0.006 |
| | (0.012) | (0.010) |
| cutcloud(40,60] | −0.052*** | 0.032*** |
| | (0.014) | (0.011) |
| cutcloud(60,80] | −0.051*** | 0.049*** |
| | (0.016) | (0.013) |
| cutcloud(80, Inf] | −0.084*** | 0.070*** |
| | (0.020) | (0.016) |
| City FE | Yes | Yes |
| Date FE | Yes | Yes |
| City:Year-Month FE | Yes | Yes |
| Observations | 67,972 | 67,972 |
| $R^2$ | 0.882 | 0.903 |
| Adjusted $R^2$ | 0.876 | 0.898 |
| Residual Std. Error | 100.819 | 78.818 |

*Note:* *p<0.1; **p<0.05; ***p<0.01
Standard errors are in parentheses and are clustered on city-yearmonth and date.

Table 5: User-Level Weather and Expressed Sentiment, Twitter No Weather Terms

| | *Dependent variable:* | |
|---|---|---|
| | Positive | Negative |
| | (1) | (2) |
| cuttmax(-Inf,0] | −0.378*** | −0.035 |
| | (0.051) | (0.038) |
| cuttmax(0,5] | −0.293*** | −0.002 |
| | (0.041) | (0.037) |
| cuttmax(5,10] | −0.189*** | 0.014 |
| | (0.032) | (0.028) |
| cuttmax(10,15] | −0.118*** | 0.010 |
| | (0.024) | (0.020) |
| cuttmax(15,20] | −0.049*** | −0.004 |
| | (0.019) | (0.014) |
| cuttmax(25,30] | −0.010 | 0.005 |
| | (0.016) | (0.014) |
| cuttmax(30,35] | −0.048** | 0.025 |
| | (0.023) | (0.019) |
| cuttmax(35, Inf] | −0.119*** | 0.121*** |
| | (0.038) | (0.032) |
| cutprcp(0,1] | −0.043*** | 0.035*** |
| | (0.012) | (0.009) |
| cutprcp(1,2] | −0.063*** | 0.063*** |
| | (0.021) | (0.018) |
| cutprcp(2,3] | −0.089*** | 0.059** |
| | (0.033) | (0.028) |
| cutprcp(3,4] | −0.193*** | 0.132*** |
| | (0.046) | (0.039) |
| cutprcp(4, Inf] | −0.235*** | 0.005 |
| | (0.062) | (0.040) |
| cuttrange(5,10] | 0.020 | −0.034** |
| | (0.019) | (0.016) |
| cuttrange(10,15] | 0.040* | −0.070*** |
| | (0.021) | (0.019) |
| cuttrange(15, Inf] | 0.063** | −0.102*** |
| | (0.027) | (0.024) |
| cuthumid(-Inf,40] | −0.019 | 0.023 |
| | (0.027) | (0.024) |
| cuthumid(60,80] | −0.002 | −0.006 |
| | (0.013) | (0.011) |
| cuthumid(80, Inf] | −0.013 | 0.020 |
| | (0.017) | (0.014) |
| cutcloud(20,40] | −0.039*** | 0.008 |
| | (0.013) | (0.011) |
| cutcloud(40,60] | −0.052*** | 0.028** |
| | (0.014) | (0.012) |
| cutcloud(60,80] | −0.054*** | 0.026* |
| | (0.016) | (0.014) |
| cutcloud(80, Inf] | −0.082*** | 0.037** |
| | (0.021) | (0.016) |
| User FE | Yes | Yes |
| Date FE | Yes | Yes |
| City:Year-Month FE | Yes | Yes |
| Observations | 79,999,498 | 79,999,498 |
| $R^2$ | 0.127 | 0.106 |
| Adjusted $R^2$ | 0.123 | 0.102 |
| Residual Std. Error | 32.779 | 26.137 |

*Note:* *p<0.1; **p<0.05; ***p<0.01
Standard errors are in parentheses and are clustered on city-yearmonth and date.

# Alternative measures of expressed sentiment

## SentiStrength

In order to classify our Twitter data with the SentiStrength sentiment algorithm, we ran their local Java version (http://sentistrength.wlv.ac.uk/index.html). SentiStrength is designed to "estimate the strength of positive and negative sentiment in short texts, even for informal language". Below we replicate our results using the SentiStrength classifier.

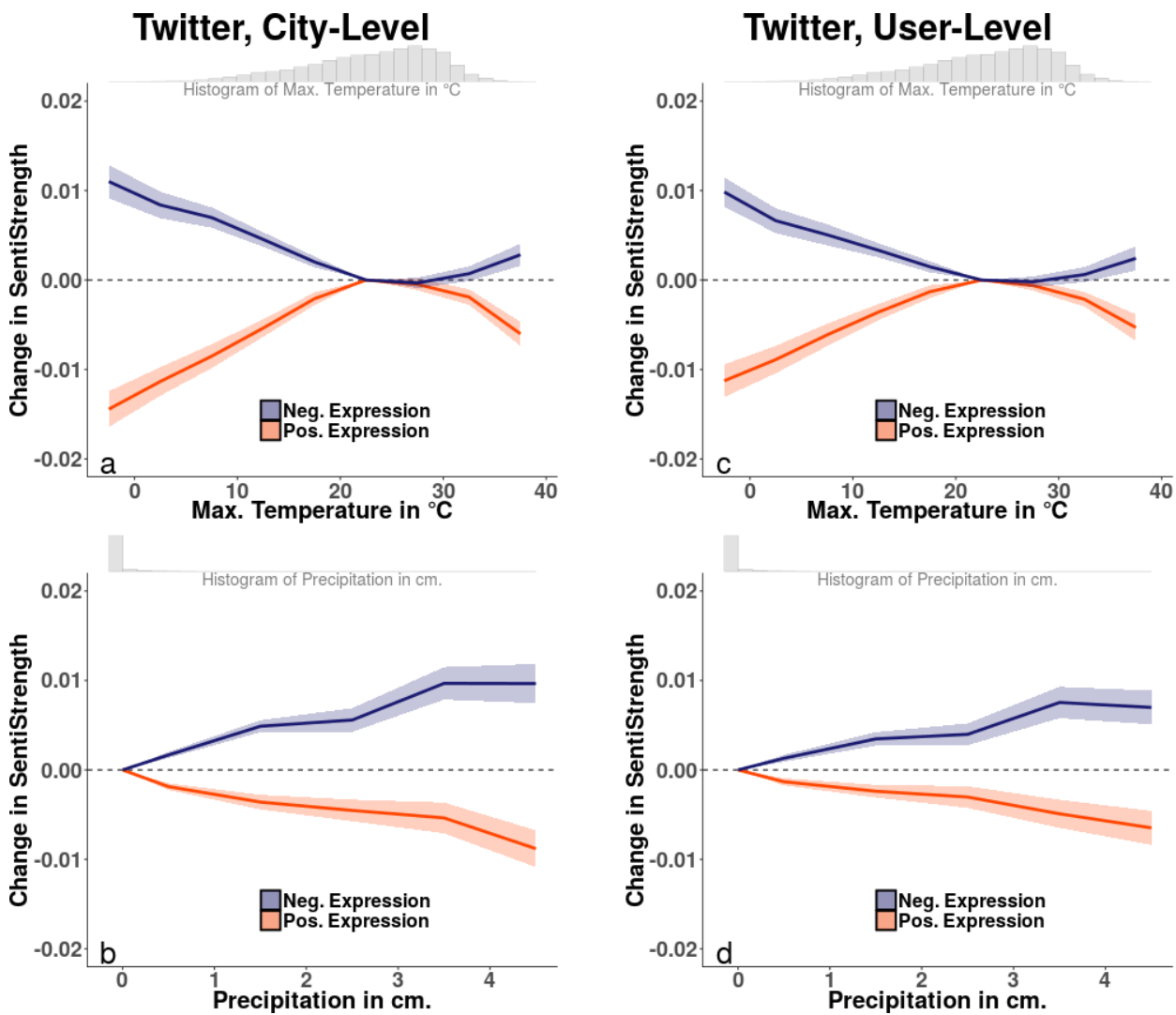

Figure 6: **Replication using SentiStrength classification.**

## Hedonometer

In order to classify our Twitter data with the Hedonometer sentiment algorithm, we employ their publicly available library (http://hedonometer.org/api.html), taking the city-day average across users' classified tweets. Hedonometer was built from a large corpus of words that were originally classified for sentiment by human workers. Below we replicate our results using the Hedonometer classifier.

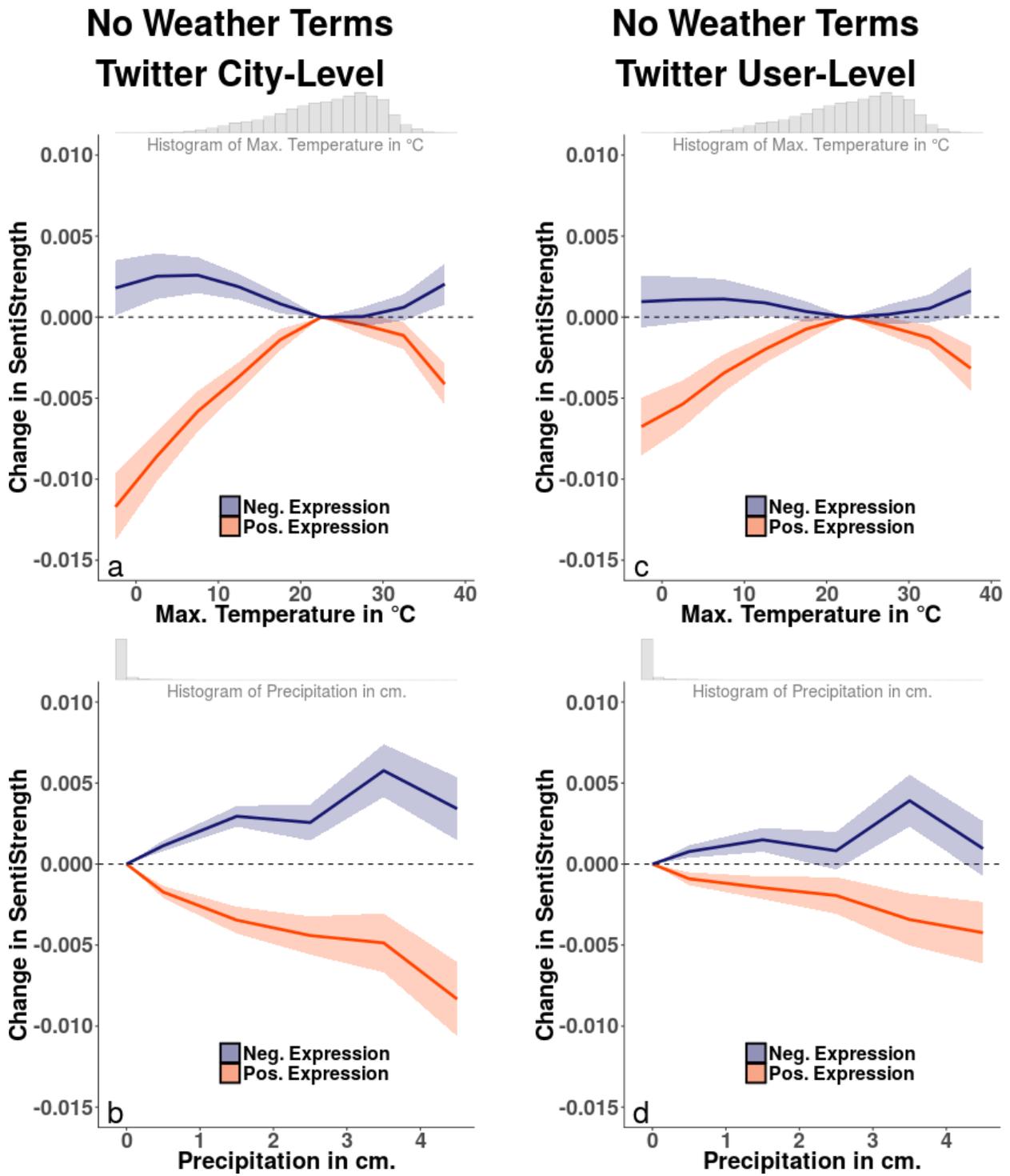

Figure 7: **Replication using SentiStrength classification.**

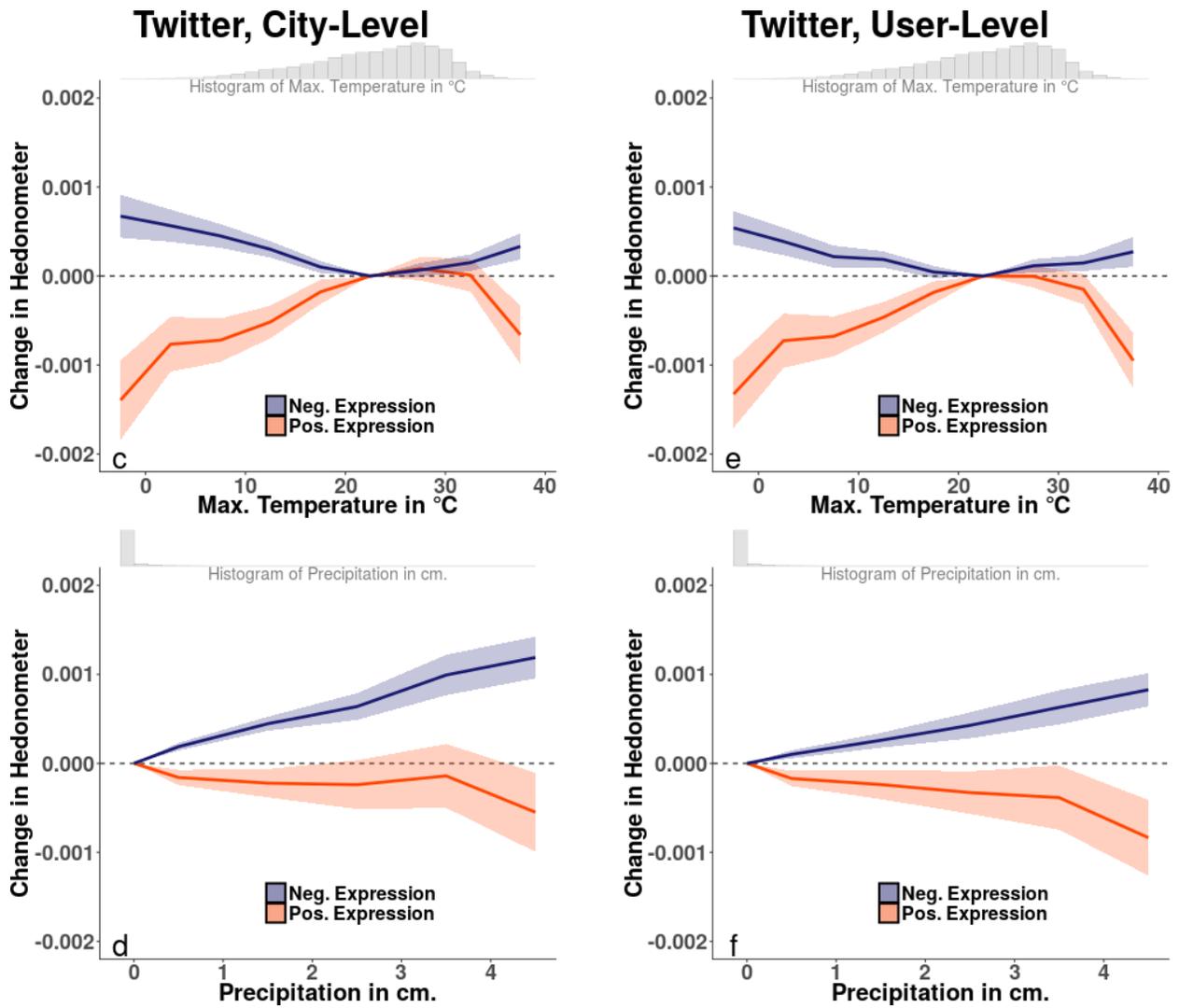

Figure 8: **Replication using Hedonometer classification.**

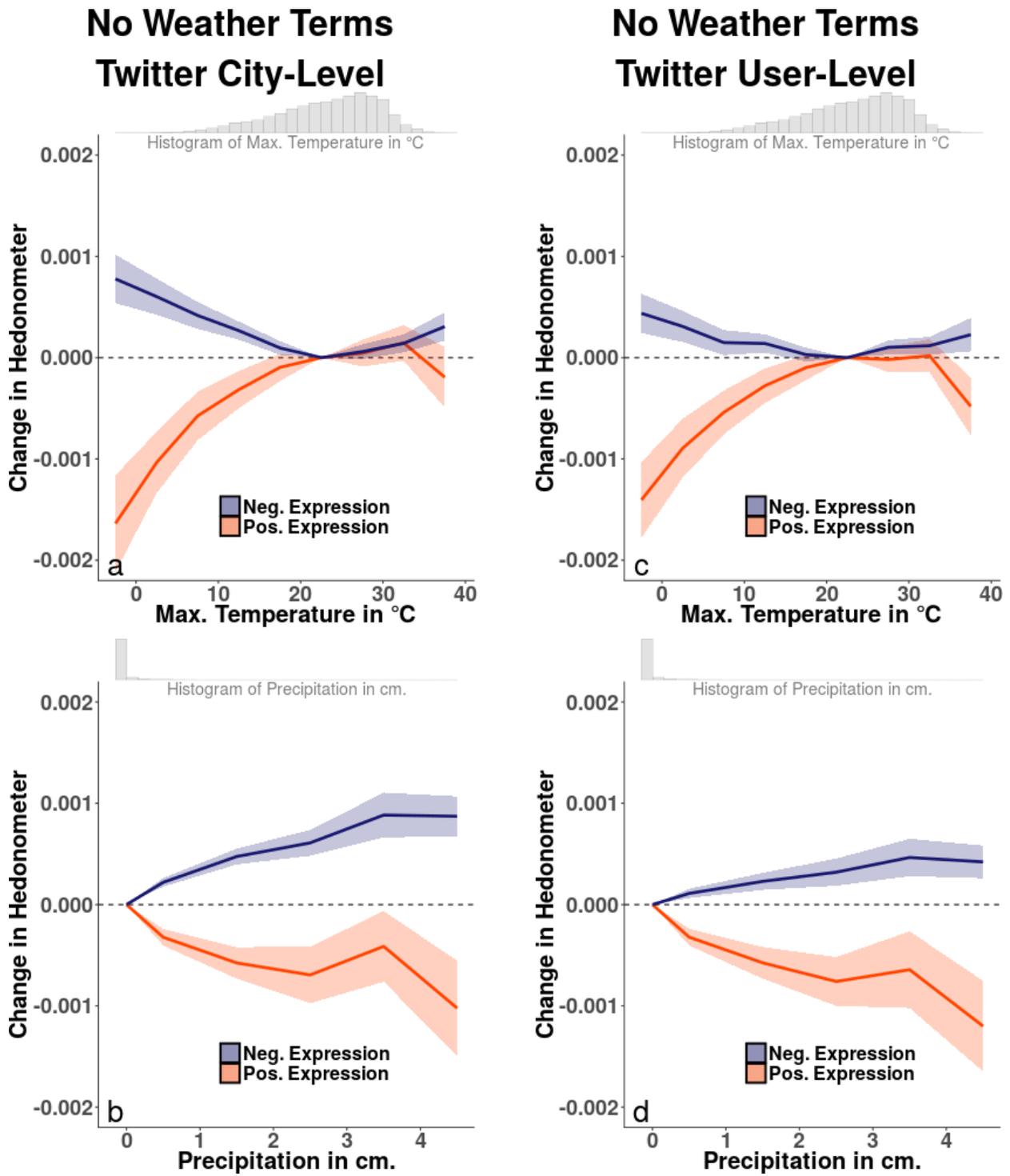

Figure 9: **Replication using Hedonometer classification.**

## Correlation of classifiers of expressed sentiment

Figure S10 displays the positive correlation observed between our positive metrics of expressed sentiment: LIWC, SentiStrength, and Hedonometer. The Hedonometer metric is moderately positively correlated with the other two, but exhibits notably less correlation than LIWC and SentiStrength exhibit between one another.

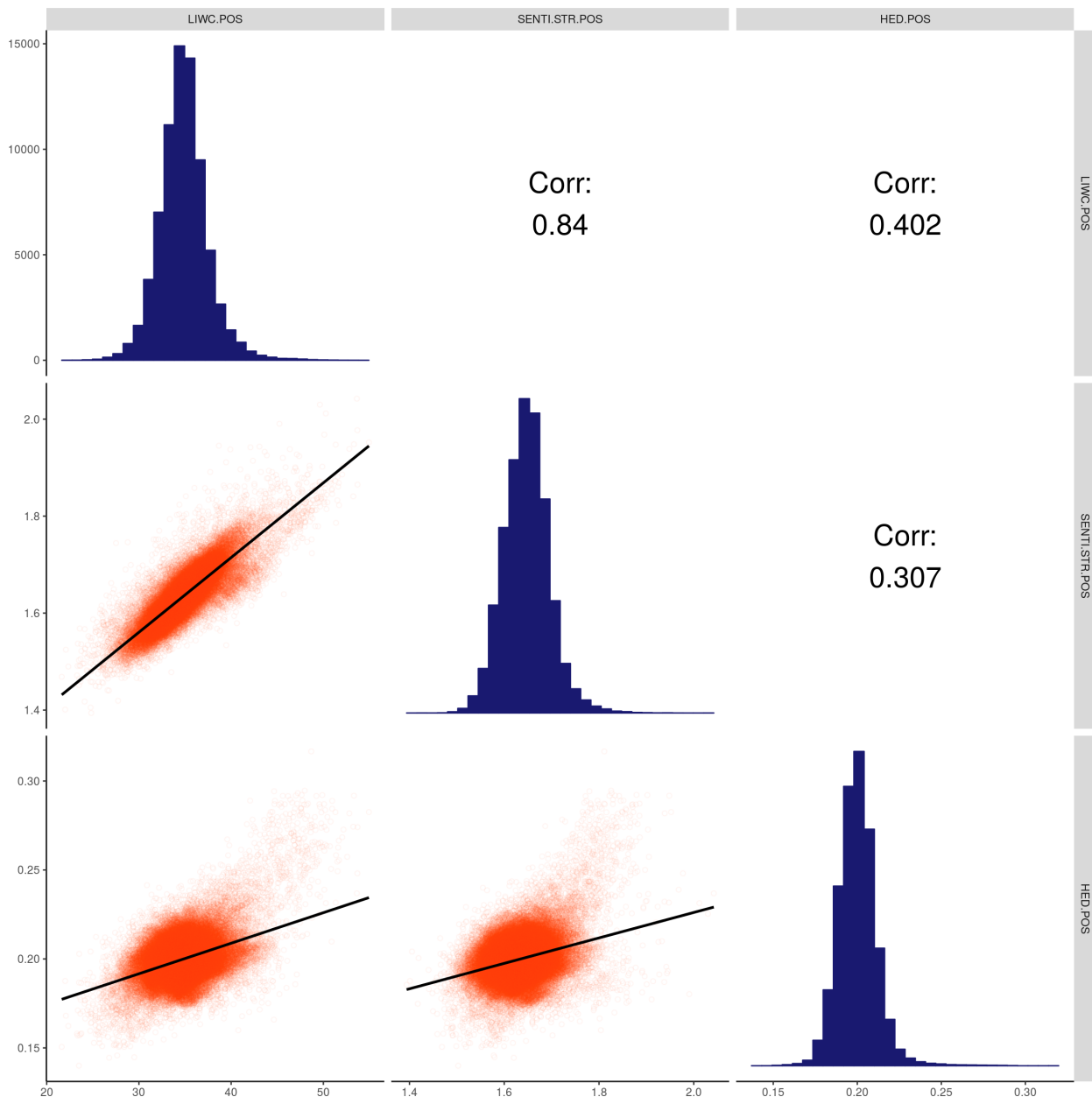

Figure 10: **Correlation between alternative positive sentiment state metrics.**

Figure S11 displays the positive correlation observed between our negative sentiment classification metrics As can be seen, all three metrics share high correlations on their classification of negative expressed sentiment.

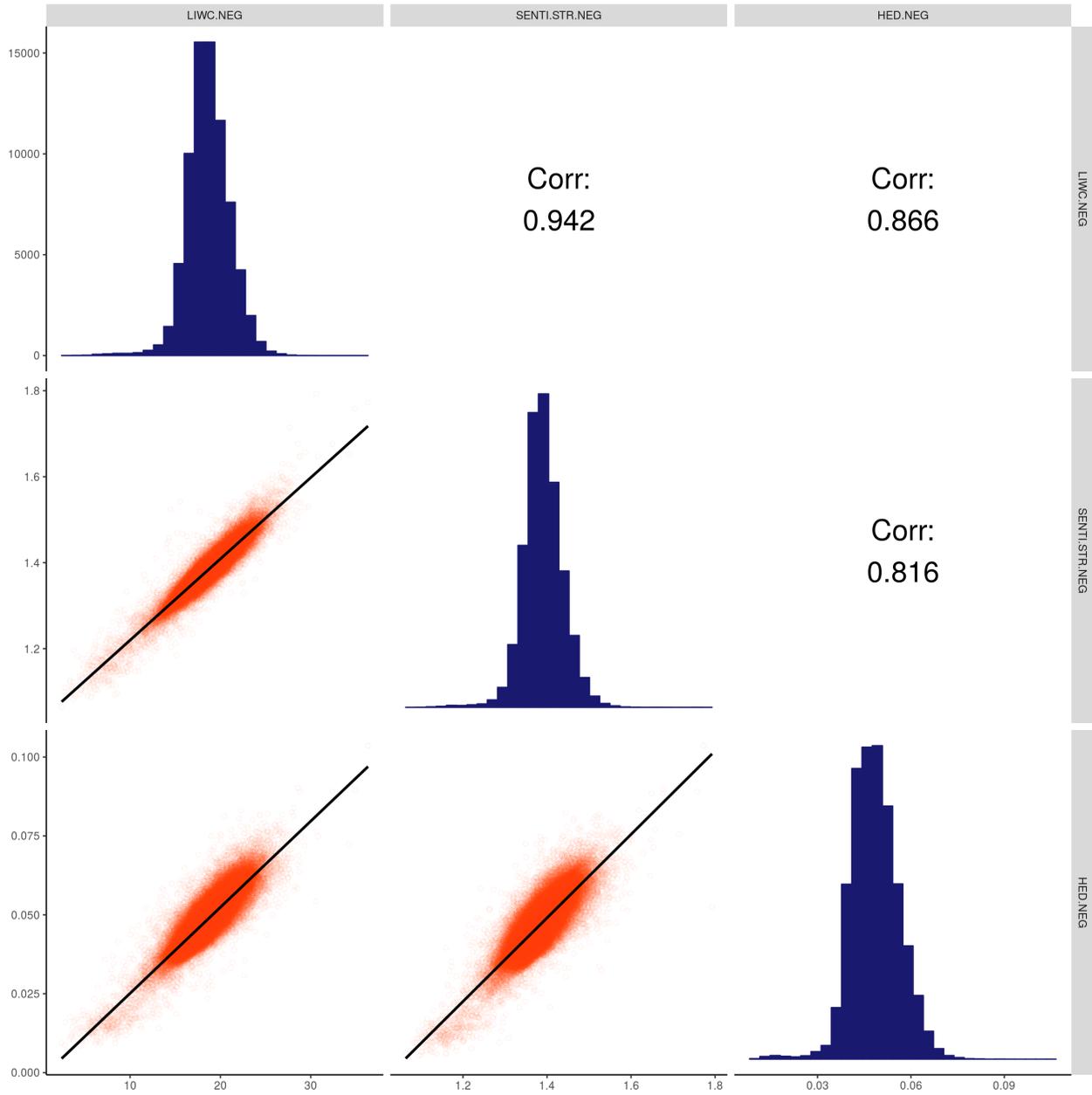

Figure 11: **Correlation between alternative negative expressed sentiment metrics.**